\newcommand{\bs}[1]{\mbox{\boldmath $#1$}}
\input psfig
\documentstyle[preprint,prb,aps]{revtex}
\begin{document}
\preprint{}
\draft
\title{Resistive magnetohydrodynamic equilibria in a torus}
\author{David Montgomery, Jason W. Bates, and H. Ralph Lewis}
\address{Department of Physics and Astronomy \\
Dartmouth College \\
Hanover, New Hampshire $ $ 03755-3528 U.S.A.}
\date{\today}
\maketitle
\begin{abstract}
It was recently demonstrated that static, resistive,
magnetohydrodynamic equilibria, in the presence of spatially-uniform
electrical conductivity, do not exist in a torus under a standard set
of assumed symmetries and boundary conditions. The difficulty, which
goes away in the ``periodic straight cylinder approximation,'' is
associated with the necessarily non-vanishing character of the curl of
the Lorentz force, $ $ ${\bf j}\bs{\times} {\bf B}$. Here, we ask if
there exists a spatial profile of electrical conductivity that permits
the existence of zero-flow, axisymmetric resistive equilibria in a
torus, and answer the question in the affirmative. However, the
physical properties of the conductivity profile are unusual (the
conductivity cannot be constant on a magnetic surface, for example)
and whether such equilibria are to be considered physically possible
remains an open question.
\end{abstract} 
\vspace{0.5in}
\pacs{PACS numbers: 52.30.Bt, 52.30.Jb, 52.55.-s, 52.65.Kj}
\narrowtext
\section{INTRODUCTION}
Discussions of the magnetic confinement of plasmas and their stability
usually begin with the subject of magnetohydrodynamic (MHD) equilibria. 
\cite{Bateman78,Wesson87} In the present generation of devices, dominated 
by the tokamak concept, the magnetofluids usually carry currents and contain
strong dc magnetic fields whose sources are in external coils. 
\par
Since the earliest days of fusion research, the various MHD equilibria have
been treated as ideal. They are assumed to have infinite electrical 
conductivity and no flow, so that Ohm's law,
\begin{equation}
{\bf E} + {\bf v} \bs{\times} {\bf B} = {\bf j}/\sigma, \label{eq: Ohm}
\end{equation}
is satisfied by taking all three terms equal to zero. Here, ${\bf E}$
is the electric field, ${\bf v}$ is the fluid velocity, ${\bf j}$ is
the electrical current density, and $\sigma$ is the electrical
conductivity, allowed to become infinite. In familiar ``Alfvenic''
dimensionless units, the curl of the magnetic field ${\bf B}$ is just
${\bf j}$, and of course, the divergences of ${\bf B}$ and ${\bf j}$
are zero. In the limit of infinite conductivity, the sharp connection
that must exist between ${\bf E}$ and ${\bf j}$ for large but finite
conductivity is, of course, destroyed. There have been attempts to
reintroduce the effects of finite conductivity ({\it e.g.}, see Grad
and Hogan \cite{Grad68}), but these attempts are in our opinion not a
satisfactory resolution; too many possibilities are raised to be able
to deal with any one of them conclusively.
\par
With the infinite conductivity assumption, the only remaining MHD
requirement for equilibrium is one of mechanical force balance, with
the local Lorentz force ${\bf j} \bs{\times} {\bf B}$ taken equal to
$\nabla p$, the mechanical pressure gradient. Such ideal equilibria
are plentiful \cite{Bateman78,Wesson87} and are not constrained by the
Ohm's law. The Grad-Shafranov formulation
\cite{Bateman78,Wesson87,Grad,Shafranov} 
then provides a framework in which axisymmetric, toroidal 
ideal equilibria can be calculated.
\par
Investigating the stability of such ideal equilibria can be a lengthy 
undertaking. This is particularly true when the program includes 
investigating the various ``resistive instabilities'' that the ideal 
equilibria may be thought heir to. Aside from the questions of consistency 
raised by ignoring resistive terms when constructing the equilibria, but 
then including them in the linear stability analysis, the possibilities
for uncovering new variants of non-ideal, normal modes growing about ideal
steady states can seem almost limitless.
\par
The purpose of this paper is to re-open the subject of zero-flow
MHD steady states by considering the implications of retaining 
Eq.~(\ref{eq: Ohm}) with finite $\sigma$. Attention will be confined to 
axisymmetric steady states with zero flow (${\bf v} = 0$) in the simplest 
toroidal geometry. A previous note by Montgomery and Shan 
\cite{Montgomery94} reported a proof of a somewhat unanticipated 
result: no such equilibria exist for the case of spatially uniform $\sigma$. 
Here, we inquire into what kinds of spatial dependences of the conductivity 
$\sigma$ will permit axisymmetric, toroidal steady states without flow. We 
do not answer the question of how such conductivity profiles might occur, or 
indeed whether they are to be expected on physical grounds. 
\par
The hydrodynamic precedent suggests the need for such inquiries, as
illustrated by plane shear flows such as pipe flow, plane Poiseuille
flow, or flow around obstacles ({\it e.g.}, see Acheson
\cite{Acheson90}). In the case of pipe flow, for example, a uniform
pressure gradient balances the axial viscous body force in the steady
state. In the ``ideal'' limit, both may be allowed to go to zero
proportionately, but the connection between them must not be lost, or
what was a well-determined (parabolic) velocity profile may become
anything at all; any axial velocity dependence upon radius is an
acceptable steady state, clearly a nonsensical conclusion.  Yet the
stability of each one of these possible profiles could be investigated
{\it ad infinitum}.
\par
It is perhaps worth remarking that in the absence of flow
velocity, the situation described here is one of classical resistive
magnetostatics, and may be treated by standard methods borrowed from
electricity and magnetism textbooks.  One can say that until a
magnetofluid begins to flow, it does not know whether it is a solid
piece of metal with a particular spatially-dependent electrical
conductivity or a plasma.  Once the requirement of balancing the
Lorentz force against a scalar pressure gradient is met, none of 
the additional approximations or assumptions of
magnetohydrodynamics {\it per se} need to be used for this problem.
\par
We also mention that recently, Bates and Lewis \cite{Bates96} have 
developed mathematical machinery for treating toroidal equilibrium profiles
in true toroidal coordinates. They were able to substantiate the result of 
Montgomery and Shan \cite{Montgomery94} in considerably more detail.
\par
It is important that the point of view adopted here be stated
clearly. We believe the full set of time-independent MHD equations,
including resistivity, velocity fields, tensor transport coefficients,
and realistic boundary conditions, are far too difficult to solve at
this moment.  This is the reason that nearly all MHD equilibrium
theory has been ideal, and has omitted these effects, even when it has
been asserted that "resistive instabilities" were being calculated. We
are interested in moving as far as we can to remove the above
restrictions, but we are also interested in doing so only by
exhibiting explicit solutions to whatever problem is undertaken.  We
are fully aware of the desirability of including all of the
complications.  Here, only one is included: finite resistivity. It is
a new result that there exist finite-resistivity toroidal MHD
equilibria without flow, which, in many ways, do not look vastly
different from ideal equilibria.  How these states that are exhibited
here relate to more general toroidal solutions with flow and the other
possible complications is a separate, and much harder, question.
\par
This paper is organized as follows. In Sec.~II, we ask and answer the
question of what kinds of spatial profiles of scalar conductivity will
permit axisymmetric toroidal equilibria in the presence of a
stationary, toroidal inductive electric field. A partial differential
equation is developed which can be thought of as replacing the
Grad-Shafranov equation. In Sec.~III, particular cases are worked out
in detail. Implications are discussed in Sec.~IV, along with questions
about possible developments and extensions of the theory.
\begin{flushleft}      
\section{THE FINITE-CONDUCTIVITY DIFFERENTIAL EQUATION}
\end{flushleft}
In toroidal geometry (in contrast to the ``straight cylinder''
approximation \cite{Shan91,Shan93,Shan94,Montgomery95}), the
steady-state Maxwell equation, $\nabla \bs{\times} {\bf E}= 0$, has
non-trivial implications for the zero-flow Ohm's law. We use 
cylindrical coordinates $(r,\varphi,z)$ whose associated unit basis
vectors are $({\bf \hat{e}}_{r},{\bf \hat{e}}_{\varphi},{\bf
\hat{e}}_{z})$. The $z$ axis is the axis of symmetry of the torus, 
and $z = 0$ is the midplane. Axisymmetry implies that the components
of each vector field are independent of the azimuthal coordinate
$\varphi$. The center line of the torus will be taken to be the circle
$r = r_0, z = 0$. The toroidal cross section will be specified by
giving some bounding curve in the $r,z$ plane that encloses this
center line. We will choose a specific shape when illustrating the
results of the formalism. The wall of the container will be idealized
as a rigid conductor coated with a thin layer of insulating
dielectric.  We do not consider the slits and slots that are necessary
in any conducting boundary to permit the driving (toroidal) electric
field to penetrate and drive the toroidal current.  
\par
The electric field ${\bf E} = E_\varphi {\bf \hat{e}}_\varphi$ will be
assumed to result from a time-proportional magnetic flux ($z$
direction) which is confined to an axisymmetric iron core whose axis
coincides with the $z$-axis and which lies entirely inside the ``hole
in the doughnut'' of the torus.  The iron core is assumed to be very
long, so the inductive electric field it produces in response to a
magnetic flux that increases linearly with time is independent of the
$z$-coordinate. Faraday's law then implies that $E_\varphi (r) = E_0
r_0/r$ only, where $E_0$ is the value of the electric field at $r =
r_0$. This is a highly simplified cartoon of a tokamak geometry, and
it is not difficult to think of ways in which it could be made more
realistic. For example, in a more refined model ${\bf E}$ might
contain the gradient of a function of both $r$ and $z$ that obeys
Laplace's equation. We do not consider here the possibility of a
poloidal current density that could result from a tensor conductivity.
\par	
The fact that ${\bf E}$ has this form, which is dictated by Maxwell's
equations, has implications for ${\bf j} = \sigma {\bf }{\bf E}$: for a
scalar conductivity $\sigma$, ${\bf j}$ must also point in the
$\varphi$ direction.  We shall assume that $\sigma$ is a function, as
yet unspecified, of both $r$ and $z$.
\par
The magnetic field consists of a toroidal part $(B_0r_0/r) {\bf
\hat{e}}_\varphi$, and a poloidal part ${\bf B}_p(r,z)$
whose curl must be ${\bf j}$:
\begin{equation}
{\bf B} = \frac{B_0r_0}{r}{\bf \hat{e}}_\varphi + {\bf B}_p(r,z).
\end{equation}					
$B_0$ is the strength of the toroidal field at $r = r_0$; it is
supported by external poloidal windings around the torus. The
source for ${\bf B}_p$ is ${\bf j}$, so that
\begin{equation}
\nabla \bs{\times} {\bf B}_p = {\bf j} = \frac{\sigma E_0 r_0}{r}\,
{\bf \hat{e}}_\varphi. \label{eq: curlB}
\end{equation}								
The boundary condition on the magnetic field is ${\bf B} \cdot {\bf
\hat{n}} = 0$, where ${\bf \hat{n}}$ is the unit normal to the surface
of the torus.  Thus the magnetic field lines will be assumed to be
confined to the torus, and we do not inquire into the necessary
external ``vertical field,'' or other current distribution required to
satisfy boundary conditions on ${\bf B}$ outside the torus.
\par 
In addition to solving Eq.~(\ref{eq: curlB}) subject to the boundary
condition, the remaining part of the problem is guaranteeing the
force balance, $\nabla p = {\bf j} \bs{\times} {\bf B}$. This can be
done by taking the divergence of this relation and solving the
resulting Poisson equation for $p$. In order for there to exist such a
solution, it is necessary that 
\begin{equation}
\nabla \bs{\times} \left({\bf j} \bs{\times} {\bf B} \right) = 
\nabla \bs{\times} \left( j_\varphi {\bf \hat{e}}_\varphi \bs{\times} 
{\bf B}_p \right) = 0.
\label{eq: curljcb}
\end{equation}		
For uniform $\sigma$, it is not possible \cite{Montgomery94} to satisfy 
Eq.~(\ref{eq: curljcb}). In the present circumstance, the question is asked: 
``What profiles of $\sigma(r,z)$ make it possible to satisfy 
Eq.~(\ref{eq: curljcb}) and then ultimately, Eq.~(\ref{eq: curlB})?'' Note 
that we are not necessarily assuming either incompressibility or uniform 
mass density.
\par	
If we derive ${\bf B}_p$ from a vector potential, ${\bf B}_p = \nabla
\bs{\times} A_\varphi(r,z){\bf \hat{e}}_\varphi$, and compute $\nabla
\bs{\times} ({\bf j} \bs{\times} {\bf B})$, we see that Eq.~(\ref{eq:
curljcb}) is equivalent to 
\begin{equation}
{\bf B}_p \cdot \nabla \left( j_\varphi/r \right) = \left[ - \frac{\partial 
A_\varphi}{\partial z} \frac{\partial}{\partial r} + \frac{1}{r} 
\frac{\partial}{\partial r} (rA_\varphi)\frac{\partial}{\partial z}\right]
\frac{j_\varphi}{r} = 0. \label{eq: longone}
\end{equation}								
\par
The general solution of Eq.~(\ref{eq: longone}) is $j_\varphi/r =
f(rA_\varphi)$, where $f$ is any differentiable function of its
argument. The only remaining equation to satisfy is Eq.~(\ref{eq:
curlB}), or $\nabla \bs{\times} \nabla \bs{\times} A_\varphi {\bf
\hat{e}}_\varphi = j_\varphi {\bf \hat{e}}_\varphi$, which can be
written as
\begin{equation}
\frac{\partial^2 A_\varphi}{\partial r^2} + \frac{1}{r} \frac{\partial A_\varphi}{\partial r} - \frac{A_\varphi}{r^2} + \frac{\partial^2 A_\varphi}{\partial z^2} = -rf(rA_\varphi). \label{eq: ellip}
\end{equation}								
Eq.~(\ref{eq: ellip}) is structurally similar to the Grad-Shafranov
equation; any choice of $f$ will lead to a partial differential
equation for $A_\varphi(r,z)$ which must be solved subject to boundary
conditions.  However, a new restriction on the spatial conductivity is
implied. Namely, the toroidal current density $rf(rA_\varphi)$ must
determine $\sigma$ by
\begin{equation}
rf(rA_\varphi) = \frac{E_0 r_0\,\sigma(r,z)}{r}\,. \label{eq: sig}		
\end{equation}						
The right hand side of Eq.~(\ref{eq: sig}) must always be positive, but 
seems otherwise unrestricted. In terms of the conventional poloidal flux 
function  $\psi = rA_\varphi$, the conductivity $\sigma$ is seen not to be 
a ``flux function.'' That is, it is not constant on a surface of constant 
$\psi$, since it is the square of $r$ times a function of $\psi$. It is 
often thought that $\sigma$ should be constant on surfaces of constant 
$\psi$, since it depends strongly on temperature and flux surfaces are often 
imagined to be surfaces of constant temperature. 
\par
For a specific form of the function $f$, the problem of finding the
solution to Eq.~(\ref{eq: ellip}) is straightforward; justifying the
spatial dependence of the conductivity implied by Eq.~(\ref{eq: sig}),
however, is not.
\par	
We pass now to the consideration of some special choices of the function 
$f$ and their consequences.
\begin{flushleft}
\section{TWO EXAMPLES}
\end{flushleft}
We do not presently inquire into the physical basis of any 
possible choice of $f(rA_\varphi )$, such as, say, a maximum-entropy or 
minimum-energy choice. Rather, we select two examples largely 
on the basis of algebraic simplicity, and also choose a toroidal 
cross section for which the equation becomes separable and the 
boundary conditions tractable. Even then, both cases exhibit some 
complexity.
\begin{flushleft}
\subsection{The choice f = constant}
\end{flushleft}
The simplest choice is to set $f$ equal to a positive constant
$\lambda > 0$, making the toroidal current density $j_\varphi$
proportional to $r$ and independent of $z$.  It should be noted that
this $r$-proportionality (for vorticity) was found useful by Norbury
\cite{Norbury73} in a study of vortex rings.  These were ideal vortex
rings, however, and required no external agency to maintain them
against dissipation. This makes the conductivity $\sigma$ vary as the
square of $r$, increasing toward the ``outboard'' side of the
torus.
\par
Eq.~(\ref{eq: ellip}), with $f = \lambda$, may be solved by first
noting that it has become a linear, inhomogeneous, partial
differential equation, the general solution to which is any particular
solution plus the most general solution of the associated homogeneous
equation. A particular solution is $A_\varphi = - \lambda r^3/8$. The
remaining homogeneous equation is the equation for a vacuum poloidal
axisymmetric magnetic field. An easy way to find this vacuum field is
to rewrite it in terms of a magnetic scalar potential $\Phi$ that
obeys Laplace's equation:
\begin{equation}
{\bf B}_p = \nabla \bs{\times} A_\varphi {\bf \hat{e}}_\varphi = - \frac{\lambda}{8} \,\nabla \bs{\times} r^3 {\bf \hat{e}}_\varphi + \nabla \Phi, \label{eq: bpoloidal}
\end{equation}	
where $\nabla^2 \Phi = 0$. The boundary condition is ${\bf B}_p \cdot
{\bf \hat{n}} = 0$ over the toroidal surface.
\par
Satisfying this boundary condition over a curved surface is a
difficult task; \cite{Montgomery94} see, however, Bates and
Lewis.\cite{Bates96} For illustrative purposes, we consider a torus
with a rectangular cross section. We assume the toroidal boundaries to
lie at $z = \pm L$, and at $r = r_{-}$ and $r = r_+$, where $r_{-}
<r<r_+$.  The vanishing of the normal component of ${\bf B}_\varphi$
at these boundaries amounts to demanding that
\begin{mathletters}
\label{eq: inhomoall}
\begin{equation}
\frac{\partial \Phi}{\partial r} = 0\,, \hskip 0.4in  {\rm at}  
\hskip 0.2in  r=r_{-},\;r_+\, 
\label{eq: boundary1} 
\end{equation}
\begin{equation}
\frac{\partial \Phi}{\partial z} =  \frac{\lambda r^2}{2}\,, 
\hskip 0.2in {\rm at} \hskip 0.2in z = \pm L\,.\;\;\;\; \label{eq: boundary2}
\end{equation}
\end{mathletters}
The general solution of $\nabla^2 \Phi = 0$ for $\Phi$ can be written as
\begin{equation}
\Phi = C_0z + \sum_k \left[C_k J_0(kr) + D_k Y_0(kr)\right] \sinh 
\left(kz + \alpha_k \right),
\end{equation}	
where $J_0$ and $Y_0$ are the usual Bessel and Weber functions,
respectively, and $C_0$, $C_k$, $D_k$, and $\alpha_k$ are arbitrary
constants. The values of $k$ remain to be determined.
\par
Eq.~(\ref{eq: boundary1}) is satisfied by requiring that 
\begin{eqnarray}
C_k J_0'(kr_-) + D_k Y_0'(kr_{-}) &=& 0\, , \label{eq: conrm} \\
C_k J_0'(kr_+) + D_k Y_0'(kr_+) &=& 0\, , \label{eq: conrp}
\end{eqnarray}	
where the primes indicate differentiation with respect to the arguments 
of the functions.  Eqs.~(\ref{eq: conrm}) and (\ref{eq: conrp}) can only be 
solved consistently for $C_k$ and $D_k$ if the determinant 
\begin{displaymath}
{\cal D} \equiv J_0'(kr_-)Y_0'(kr_+) - J_0'(kr_+)Y_0'(kr_-)
\end{displaymath}	
vanishes. For an infinite sequence of $k$-values, with each $k$
corresponding to a particular zero of ${\cal D}$ for given values of
$r_-$ and $r_+$, general Sturm-Liouville theory tells us that the
functions
\begin{equation}
\phi_k \equiv \epsilon_k \left[J_0(kr) + (D_k/C_k)\,Y_0(kr)\right],
\end{equation}	
form a complete orthonormal set on the interval $r_-<r<r_+$.  The 
$\epsilon_k$ are real constants chosen to normalize the $\phi_k$:
\begin{equation}
\int_{r_-}^{r_+} \phi_{k_1} \phi_{k_2} r\,dr = \delta_{k_1,\,k_2}\,.
\end{equation}								
\par
The $z$-boundary conditions can both be satisfied by choosing the 
$\alpha_k = 0$ for all $k$. The requirement is
\begin{equation}
\tilde{C}_0\,\phi_0 + \sum_k k\,\tilde{C}_k\,\phi_k(r)\,\cosh kL = 
\frac{\lambda r^2}{2}\,,
\end{equation}
where $\phi_0 = \epsilon_0$ is a constant. This can be achieved if the 
expansion coefficients $\tilde{C}_k$ are chosen according to
\begin{equation}
\tilde{C}_0 = \frac{\lambda \epsilon_0}{2} \int_{r_-}^{r_+}r^3\,dr\,, 
\label{eq: Ckeq0}	
\end{equation}					
for $k=0$, and
\begin{equation}
\tilde{C}_k = \frac{\lambda}{2} \int_{r_-}^{r_+} \phi_k(r) r^3\,dr/ 
\label{eq: Ckneq0}
\left[k \cosh kL\right],
\end{equation}
for $k \neq 0$. The full solution for ${\bf B}_p$ is then given by
Eq.~(\ref{eq: bpoloidal}), with
\begin{equation}
\Phi = \tilde{C}_0\,\phi_0\,z + \sum_k \tilde{C}_k\,\phi_k(r)\,
\sinh (kz)\,.		
\end{equation}					
\par
In order to determine the allowed values of $k$ in this problem, the
zeros of the determinant ${\cal D}$ must be found. The function ${\cal
D}$ is an oscillating function of $k$ that intersects the positive
$k$-axis an infinite number of times. Using standard numerical
techniques, the intersections can be computed for specified values of
$r_-$ and $r_+$, and in this way the discrete spectrum of the
permitted $k$ values determined.  For each $k$, we may calculate
$C_k/D_k$. The results for the $k$ values, the $C_k/D_k$ ratios, and
the $\epsilon_k$ normalization constants can be stored numerically,
and the expansion coefficients $\tilde{C}_k$ determined from
Eqs.~(\ref{eq: Ckeq0}) and (\ref{eq: Ckneq0}). We may then plot
magnetic surface contours, or surfaces of constant $\psi=
rA_\varphi$. The surfaces of constant pressure (since the current is
purely toroidal) are guaranteed to coincide with the $\psi$-surfaces.
\par	
To see that contours of constant pressure coincide with constant
$\psi$-surfaces, it is useful to express the poloidal magnetic field
${\bf B}_p$ in terms of $\psi$:
\begin{equation}
{\bf B}_p = \nabla \psi \bs{\times} \nabla \varphi. \label{eq: nabcross}
\end{equation}
Then, the equation for scalar-pressure equilibrium, $\nabla p = {\bf j}
\bs{\times} {\bf B}$, with ${\bf j}= \lambda r \,
{\bf \hat{e}}_\varphi$, can be integrated to give $p = \lambda \psi +
const. $ In Fig.~1, we illustrate surfaces of constant $\psi$, for
$r_-/r_0 = 0.6$, $r_+/r_0 = 1.4$ and $L/r_0 =0.3$. The current
contours, in this case, will be rather strange, since the current is
simply constant on lines of constant r, inside the torus, right up to
the boundary.
\begin{flushleft}
\subsection{The case of linear variation}
\end{flushleft}
The other case we wish to consider is that of a linear variation 
of $f$, or $j$ proportional to the square of $r$ times $A_\varphi$. The 
resulting linear differential equation is now homogeneous, which results in an 
interesting, but imperfectly understood ``quantum'' phenomenon: a preference 
for certain ratios of width to height for the rectangular cross section, 
in the steady states.
\par
It is convenient for this case to re-cast the differential equation 
in terms of the magnetic flux function $\psi=rA_\varphi$:
\begin{displaymath}
r \frac{\partial}{\partial r} \left(\frac{1}{r} \frac{
\partial \psi}{\partial r}\right) + \frac{\partial^2 \psi}{\partial z^2} =
-rj_\varphi. 
\end{displaymath}
We assume that the magnetofluid fills a torus with a rectangular 
cross section bounded by the planes $z = \pm L$, and the radii $r_- = r_0 - 
\delta$ and $r_+ = r_0 + \delta$. Because the wall of the torus is assumed to 
be perfectly conducting, we impose the boundary condition that the normal 
component of the magnetic field vanish at the wall.
\par
For $j_\varphi = \lambda r\,\psi$, the equation 
for the flux function becomes
\begin{equation}
r \frac{\partial}{\partial r} \left(\frac{1}{r} \frac{\partial \psi}{\partial 
r}\right) + \frac{\partial^2 \psi}{\partial z^2} + \lambda r^2 \psi =0. 
\label{eq: eqrz}
\end{equation} 
This equation can be solved in terms of confluent hypergeometric functions. 
To see this, we proceed as follows. First, we make
the variable substitution $\rho = \sqrt{\lambda}\,r^2/2$, which transforms 
Eq.~(\ref{eq: eqrz}) into
\begin{displaymath}
2\sqrt{\lambda}\rho\,\frac{\partial^2 \psi}{\partial \rho^2} + \frac{
\partial^2 \psi}{\partial z^2} + 2\sqrt{\lambda}\rho\,\psi =0.
\end{displaymath}
Then, seeking separable solutions of the form $\psi(\rho,z) = R(\rho)Z(z)$, 
we find
\begin{eqnarray}
\frac{d^2 Z}{dz^2} - k^2 Z = 0, \\
\frac{d^2 R}{d \rho^2} + \left(1 + \frac{k^2}{2 \sqrt{\lambda}\rho}\right)
R=0. \label{eq: confluent}
\end{eqnarray}
The equation for $Z$ results in trigonometric or hyperbolic functions, 
depending on whether $k$ is imaginary or real. Since the condition ${\bf B}_p 
\cdot {\bf \hat{n}} = 0$ on the rectangular wall of the torus leads to 
homogeneous boundary conditions on $\psi$,
\begin{mathletters}
\label{eq: homoall}
\begin{equation}
\frac{\partial \psi}{\partial r} = 0, \hskip 0.2in {\rm at} 
\hskip 0.2in z = \pm L\,\;\;\;\; \label{eq: bc1}
\end{equation}
\begin{equation}
\frac{\partial \psi}{\partial z} = 0, \hskip 0.2in {\rm at}
\hskip 0.2in r = r_+, r_-\, \label{eq: bc2}
\end{equation}
\end{mathletters}
the parameter $k$ must be imaginary. Thus, the solution for $Z$ is either 
$\sin(\kappa z)$ or $\cos (\kappa z)$, where $k = i\kappa$, and $\kappa$ is 
real. The boundary condition in Eq.~(\ref{eq: bc1}) can be fulfilled if we 
choose $Z = \cos (\kappa z)$, and require $\kappa = \pi(n + \frac{1}{2})/L$, 
where $n$ is an integer.
\par			
The solution of Eq.~(\ref{eq: confluent}) is given by
\begin{equation}
R = C_\kappa Re\left[\rho e^{-i\rho} M(1-i\eta,2;2i\rho)\right] + D_\kappa 
Re\left[\rho e^{-i\rho} U(1-i\eta,2;2i\rho)\right] ,
\end{equation}
where $C_\kappa$ and $D_\kappa$ are real constants, and $\eta \equiv
\kappa^2/4 \sqrt{\lambda}$. $M(a,c;x)$ and $U(a,c;x)$ are the regular and 
irregular, confluent hypergeometric (Kummer) functions, respectively.
\cite{Arfken85} They satisfy the second-order, ordinary
differential equation
\begin{displaymath}
x \frac{d^2 y}{dx^2} + (c-x)\frac{dy}{dx} - ay = 0.
\end{displaymath} 
\par
To fulfill the second boundary condition given in Eq.~(\ref{eq: bc2}), 
we demand
\begin{eqnarray}
C_\kappa Re\left[ e^{-i\rho_-} M(1-i\eta,2;2i \rho_-)\right] + D_\kappa 
Re\left[ e^{-i\rho_-}U(1-i\eta,2;2i \rho_-)\right] = 0, \label{eq: conda} \\
C_\kappa Re\left[ e^{-i\rho_+}M(1-i\eta,2;2i \rho_+)\right] + D_\kappa 
Re\left[ e^{-i\rho_+}U(1-i\eta,2;2i \rho_+)\right] = 0, \label{eq: condb}
\end{eqnarray}
where $\rho_{\pm} \equiv \sqrt{\lambda}(r_0 \pm
\delta)^2/2$. Eqs.~(\ref{eq: conda}) and (\ref{eq: condb}) only have a
solution for $C_\kappa$ and $D_\kappa$ if the determinant vanishes,
{\it i.e.},
\begin{eqnarray}
& &Re\left[e^{-i\rho_-}M(1-i\eta,2;2i \rho_-)\right]\,\times Re\left[e^{-i\rho_+}U(1-i\eta,2;2i
\rho_+)\right] \nonumber \\
\mbox{} & & - Re\left[e^{-i\rho_+}M(1-i\eta,2;2i \rho_+)\right]\,
\times Re\left[e^{-i\rho_-}U(1-i\eta,2;2i \rho_-)\right] = 0. \label{eq: deterMU}
\end{eqnarray}
For each integer $n$, this equation holds only for a limited
combination of $\lambda$, $r_0$, $\delta$, and $L$. That is,
for a particular value of the current density, only certain aspect
ratios of the rectangular toroidal wall are allowed if steady state
solutions are to exist.
\par
There is an additional physical constraint to be considered here which
limits the values that the parameters $n$, $\lambda$, $r_0$, $\delta$,
and $L$ can assume: the current density $j_\varphi$ must not change
sign. Consequently, it seems that the only permissible value for $n$
is $n=0$. Note that the homogeneity of Eq.~(\ref{eq: eqrz}) has
eliminated the need for an eigenfunction expansion and satisfaction of
boundary conditions, as in Eqs.~(\ref{eq: Ckeq0}) and (\ref{eq:
Ckneq0}). We have not experimented with the possibilities of linear
combinations of the two types of $f$ used in this section, but the
range of possibilities is clearly wide.
\par
In the case that $f$ varies linearly with its argument, the pressure
is given by $p = \lambda\,\psi^2/2 + const.$, which is easily obtained
from integrating $\nabla p = {\bf j} \bs{\times} {\bf B}$, with ${\bf
j} = \lambda r \,\psi \,\bf{\hat{e}}_\varphi$ and ${\bf B}_p$ written
in the form of Eq.~(\ref{eq: nabcross}). Thus, once again, the
pressure does not vary on surfaces of constant $\psi$ (magnetic
surfaces).
\par
For $n=0$, one set of parameters that satisfies Eq.~(\ref{eq:
 deterMU}) is $\lambda\,r^4_0=29.375$, $\delta/r_0=0.4$, and
 $L/r_0=0.4$. Contour plots of $\psi$, $j_\varphi$, and $\sigma$ using
 these values appear in Fig.~2. One can see, from Fig.~2, that no
 radical qualitative departures from the magnetic surfaces expected
 from a Grad-Shafranov treatment have been found; the ``Shafranov
 shift'' of the magnetic axis to the outboard side of the torus is
 clearly evident.  The principal difference is the non-coincidence of
 the surfaces of constant conductivity and the magnetic surfaces.
\\
\section{DISCUSSION AND CONCLUSIONS}
In this paper, we have provided a framework in which toroidal
axisymmetric, resistive steady states can been constructed for the MHD
equations with scalar conductivity.  Our approach is to search for a
conductivity profile that permits such steady states to exist.
Admittedly, the result is artificial since no discussion has been
given of how such a profile might arise, or be consistent with an
energy equation.  If ideal MHD had not dominated the subject of
magnetic confinement theory for forty years, the exercise might be
considered unmotivated. However, since the proof of the nonexistence
of toroidal resistive steady states with uniform
conductivity,\cite{Montgomery94,Bates96} the question has arisen
whether {\em any} current profile will support a static MHD state;
here, we have answered this question affirmatively. We consider the
formalism for constructing static toroidal resistive states,
incomplete as it is, to be physically less objectionable than an
ideal treatment, which in fact still underlies the vast
majority of instability calculations.
\par
What actually seems more likely, based on earlier dynamical
computations\cite{Shan93,Montgomery95,Dahlburg88} using the full
three-dimensional MHD equations, is that {\em velocity fields} permit
Ohm's law to be satisfied in resistive steady states of confined
magnetofluids. Even if this conjecture is correct, though, the effects
of finite flow will be bound up with that of the conductivity, which
is virtually guaranteed to be spatially non-uniform. The satisfaction
of the poloidal and toroidal components of both Ohm's law and the
equation of motion, simultaneously, is an arduous task when velocity
fields are included; at present, no one seems close to solving this
problem. The situation becomes even more difficult if one demands that
the pressure be derived from a local equation of state. In this case,
complete consistency demands the simultaneous satisfaction of an
energy equation as well -- a formidably difficult undertaking ({\it
e.g.}, see Goodman\cite{Goodman93}).
\par
We should remark that the equilibria exhibited here could have been
obtained formally as Grad-Shafranov equilibria with a constant poloidal
current function, and with a proper choice of the pressure function.  A
demand for consistency with Ohm's law leads again to a determination
of the conductivity through Eq.~(\ref{eq: sig}), with $rA_\phi$
replaced by $\psi$, and $f$ replaced by the derivative of the pressure
function with respect to its argument ($\psi$).
\par
We should remark on a perception of several decades ago, due to
Pfirsch and Schl\"{u}ter,\cite{Pfirsch62} that difficulties associated
with retaining Ohm's law and finite resistivity led to difficulties
for ideal MHD equilibrium theory. Their resolution, though never
absorbed to any significant degree in working models for tokamak
equilibria, was to attempt to satisfy the Ohm's law in perturbation
theory by using it to calculate iteratively the two perpendicular
components of a fluid velocity to be associated with any given ideal
equilibrium.  A velocity field was thus taken into account in the
Ohm's law, but not in the equation of motion. A perturbation expansion
in inverse aspect ratio was also featured.  The conclusion was reached
that a universal plasma outflow existed (sometimes loosely called a
``diffusion velocity'') that had to be compensated by ``sources'' of new
plasma within the plasma interior (never identified quantitatively).
We cannot accept this conclusion.  The explicit examples shown here,
which do not rely on any perturbation expansions or other
approximations, demonstrate the existence of zero-flow non-ideal
solutions without plasma losses, and as such explicitly contradict the
Pfirsch-Schl\"{u}ter expressions for positive-definite outward flux. We do
believe, however, that Pfirsch and Schl\"{u}ter were correct in their
assumption that real-life toroidal MHD steady states will involve mass
flows in a fundamental way.  It is in a sense remarkable that there
has been so little attention to this as yet unresolved problem.
\par
The main point seems to us to be the need for developing a renewed
respect for the problem of determining allowed steady states of a
plasma.  Neither present diagnostics nor present theoretical machinery
permit it. The phasing out of the vocabulary that has arisen in
connection with ideal steady states will require the passage of some
time.
\\
\acknowledgments
One of us (DM) wishes to thank Professor Xungang Shi for a helpful
discussion of vortex rings. This work was supported in part by the
Burgers Professorship at Eindhoven Technical University in the
Netherlands, and at Dartmouth by a Gordon F. Hull Fellowship, and a
U.S.~Dept.~of Energy Grant DE-FGO2-85ER53194.

\begin{figure} 
%\centerline{\psfig{figure=DCM_EPS_Figures/dcm_jwb_hrl_Ffig1.eps}}
\centerline{\psfig{figure=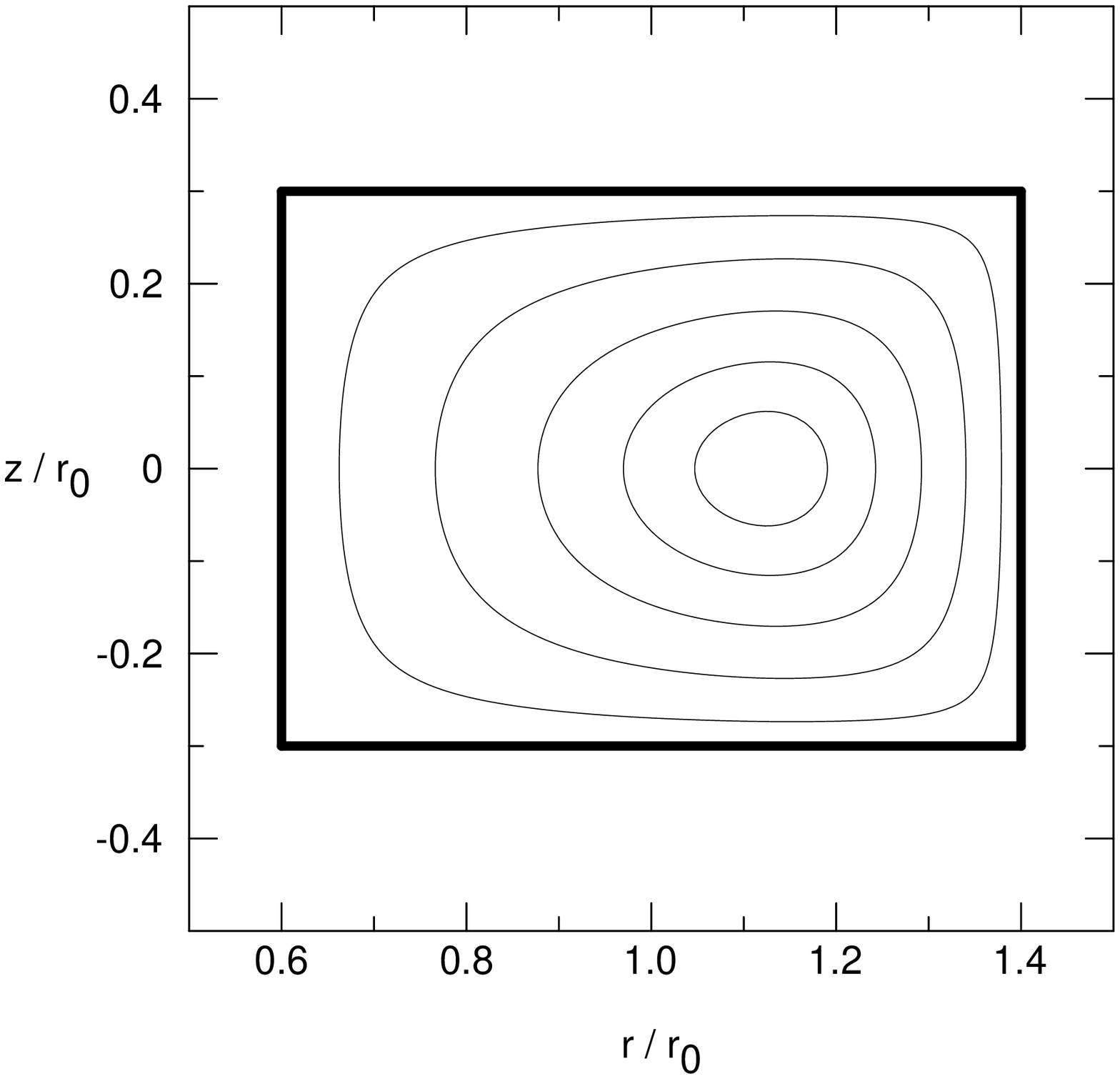}}
\caption{Contours of constant $\psi$ in the case that $j_\varphi=\lambda r$, 
for $r_-/r_0=0.6$, $r_+/r_0=1.4$ and $L/r_0=0.3$. The ratio of magnetofluid
pressure to poloidal magnetic field pressure ($\beta_p$) in this case is 
about 1.1.}
\label{fig1}
\end{figure}
\begin{figure} 
%\centerline{\psfig{figure=DCM_EPS_Figures/dcm_jwb_hrl_Ffig2a.eps}}
%\centerline{\psfig{figure=DCM_EPS_Figures/dcm_jwb_hrl_Ffig2b.eps}}
%\centerline{\psfig{figure=DCM_EPS_Figures/dcm_jwb_hrl_Ffig2c.eps}}
\centerline{\psfig{figure=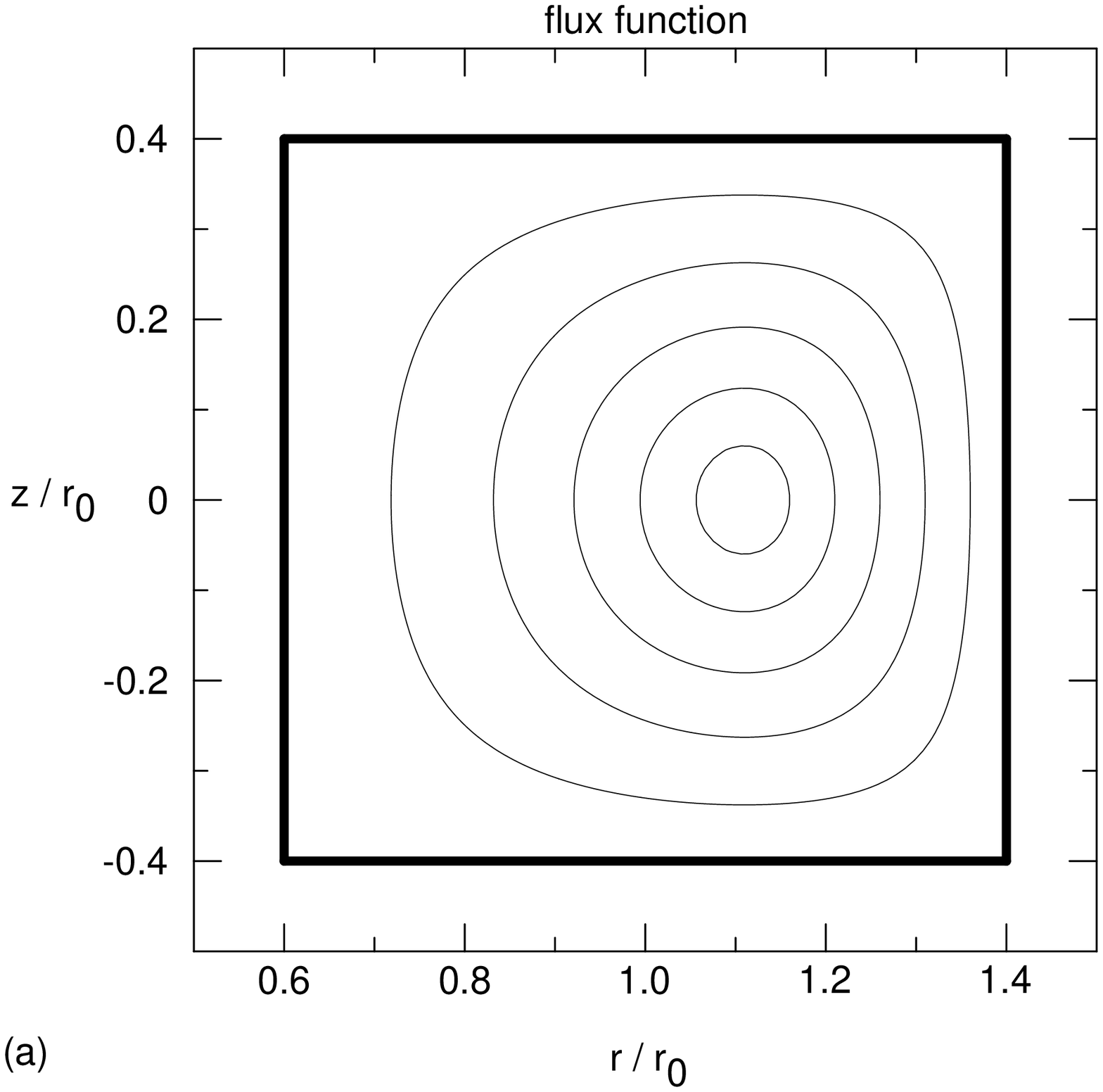}}
\centerline{\psfig{figure=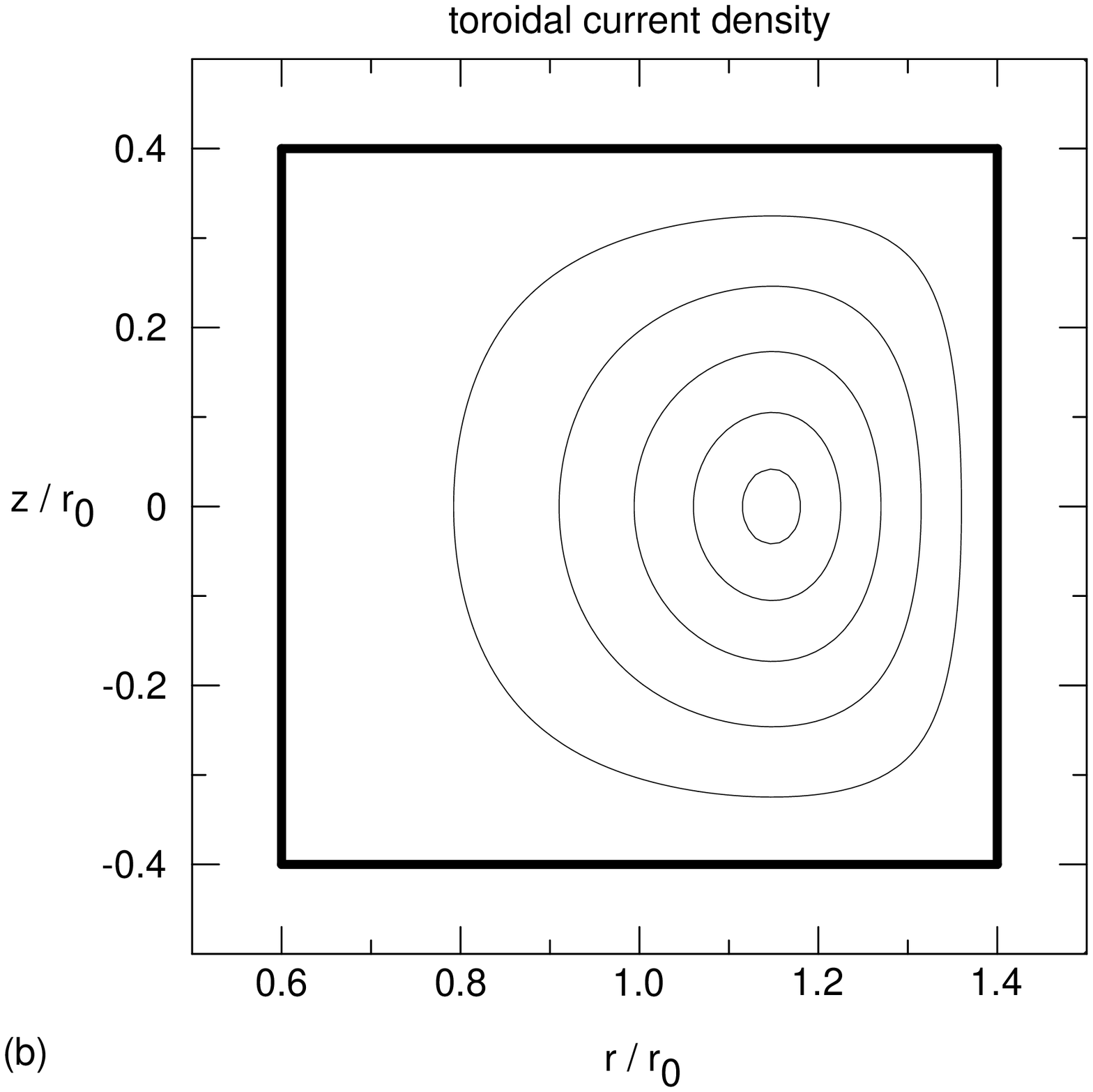}}
\centerline{\psfig{figure=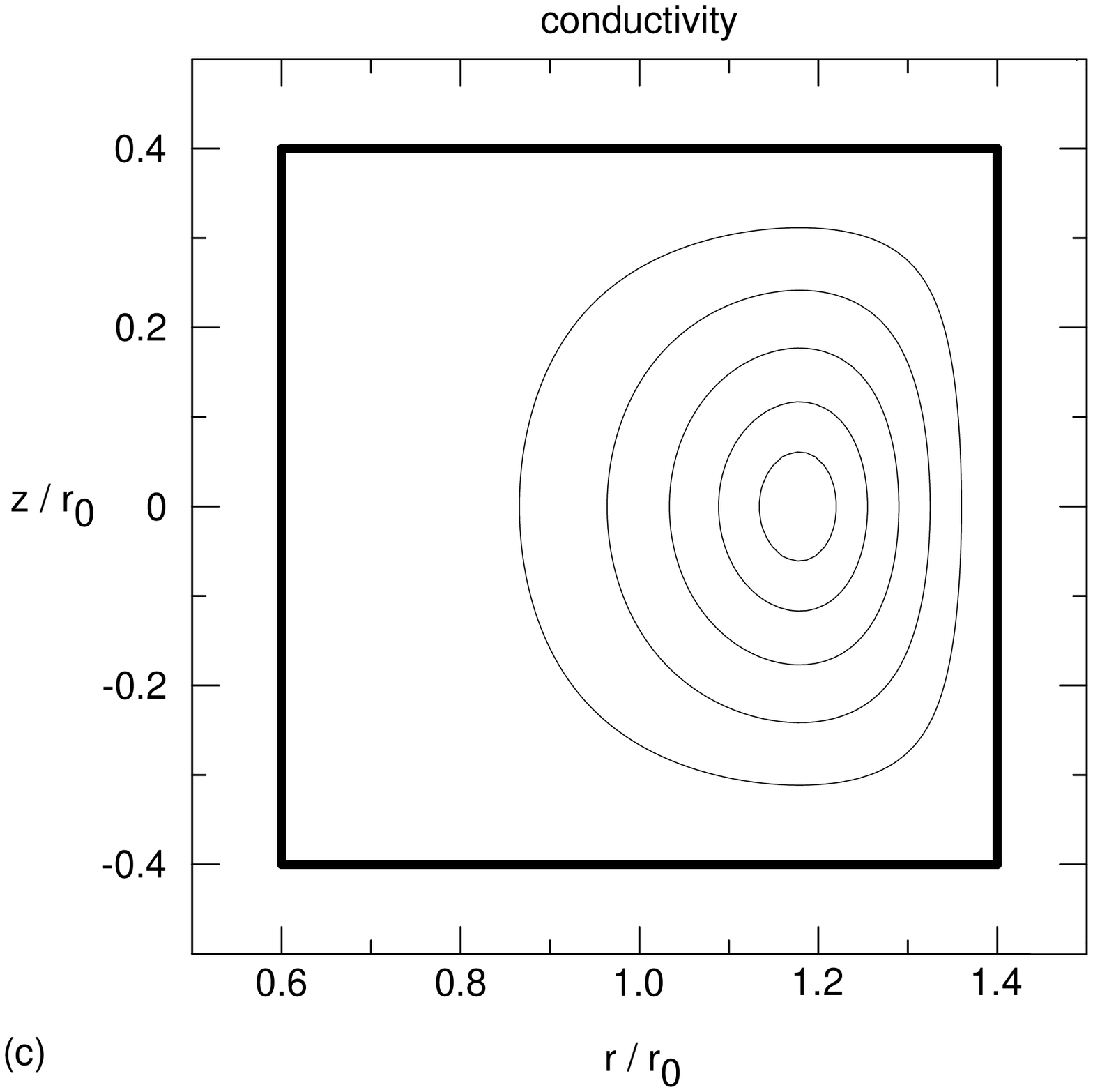}}
\caption{ (a) Contours of constant $\psi$ (and pressure) in the case that 
$j_\varphi=\lambda r\,\psi$, for $n=0$, $\lambda\,r^4_0=29.375$,
$\delta/r_0=0.4$, and $L/r_0=0.4$. (b) Current density contours, and
(c) conductivity contours for the same set of parameters. The square
box shows the location of the perfectly conducting, toroidal
boundary. Note the non-coincidence of the conductivity contours and
the magnetic surfaces. In this example, $\beta_p \simeq 1.2$.}
\label{fig2}
\end{figure}
%\begin{figure} 
%\centerline{}
%\caption{Current density contours for $\lambda\,r^4_0=1$, $\delta/r_0=0.4$, 
%and $L/r_0=0.403$.}
%\label{fig3}
%\end{figure}
%\begin{figure} 
% \centerline{}
%\caption{}
%\label{fig4}
%\end{figure}
\end{document}